\documentclass[
prd,aps,
showpacs,tightenlines,nofootinbib,preprintnumbers, superscriptaddress
]{revtex4}
\usepackage{graphicx}
\usepackage{amssymb,amsmath,latexsym}
\usepackage{amsfonts}

\usepackage{cancel}
\usepackage{color}
\input{colordvi.tex}
\input{epsf}
\usepackage{epsf}
\usepackage{graphicx,epsfig}
\usepackage{bm}
\usepackage{latexsym,float}

\newcommand{\Om}{\Omega}

\newcommand{\bear}{\begin{array}}  \newcommand{\eear}{\end{array}}
\newcommand{\beq}{\begin{equation}}  \newcommand{\eeq}{\end{equation}}
\newcommand{\bef}{\begin{figure}}  \newcommand{\eef}{\end{figure}}
\newcommand{\bec}{\begin{center}}  \newcommand{\eec}{\end{center}}

\newcommand{\order}{{\cal O}}

\newcommand{\beqa}{\begin{eqnarray}}
\newcommand{\eeqa}{\end{eqnarray}}
\newcommand{\p}{\phi}

\newcommand{\ka}{\kappa}
\newcommand{\Omp}{\Omega_{\phi}}
\newcommand{\wzero}{w_{(0)}}
\newcommand{\Gammazero}{\Gamma_{(0)}}
\newcommand{\phizero}{\phi_{(0)}}

\newcommand {\ga} {\ {\raise-.5ex\hbox{$\buildrel>\over\sim$}}\ }
\newcommand {\la} {\ {\raise-.5ex\hbox{$\buildrel<\over\sim$}}\ } 
\newcommand{\eqn}[1] {Eq.~(\ref{#1})}

\def\be{\begin{equation}}
\def\ee{\end{equation}}
\def\ba{\begin{eqnarray}}
\def\ea{\end{eqnarray}}

\begin{document}

\title{The Equation of State of Tracker Fields}

\author{Takeshi Chiba}
\affiliation{Department of Physics, \\
College of Humanities and Sciences, \\
Nihon University, \\
Tokyo 156-8550, Japan}

\date{\today}


\begin{abstract}
We derive the equation of state of tracker fields, which 
are typical examples of freezing quintessence (quintessence with 
the equation of state approaching toward $-1$), taking into account of 
the late-time departure from the tracker solution due to the nonzero density 
parameter of dark energy $\Omp$. We calculate the equation of state as a function of 
$\Omp$ for constant $\Gamma=VV''/(V')^2$ (during matter era) models. 
The derived equation of state  
contains a single parameter, $w_{(0)}$, which parametrizes the equation of state during 
the matter-dominated epoch. We derive observational constraints on $w_{(0)}$ and 
find that observational data are consistent with the cosmological constant: 
$-1.11< \wzero< -0.96 (1 \sigma)$. 
\end{abstract}

\maketitle

\section{Introduction}

There is strong evidence that the Universe is dominated by dark energy. 
Moreover, the cosmological constant fits the current observational data quite well. 
However, how much is a dark energy model close to the cosmological constant? 
In order to quantify such "distance from the cosmological constant" in the 
dark energy theory space, we need to introduce a parametrization of the equation of state, 
$w(a)$,  which parametrizes the deviation from the cosmological constant, $w=-1$. 

In our former study \cite{ds,chiba,cds}, we derived the equation of state for
 certain scalar field dark energy (quintessence \cite{quint}/ k-essence \cite{kess}) models 
under the assumption such that the scalar field slow rolls ($w\simeq -1$) 
during the matter dominated era.  Such quintessence exhibits thawing behavior \cite{cl}:  
the scalar field freezes during the matter era and gradually moves after the dark energy dominated era 
so that the equation of state deviates from $w\simeq -1$. 
We find our parametrization applies both to thawing quintessence models and 
to a subset of thawing k-essence models with $w\simeq -1$. 

However, there are quintessence models which evolve the opposite: freezing models \cite{cl} 
with $w$ approaching toward $w=-1$ so that the scalar field gradually freezes its motion. 
In this paper, we derive the equation of state for a class of freezing quintessence models 
called tracker fields \cite{swz} whose equation state is nearly constant during 
the matter-dominated era. For freezing models, the equation of state deviates from $-1$ during the matter era 
so that the slow-roll approximation is not a good approximation. Instead, we solve the equation of 
motion by expanding around the tracker solution (the solution to which the tracker field converges 
from various initial conditions). The equation of state up to first order in the density 
parameter of dark energy $\Omp$ was derived by \cite{ws} for an inverse power-law potential \cite{power1}. 
We extend the solution to higher orders in $\Omp$ and present a useful approximation to the equation 
of state for tracker fields. Hence we now have a physically motivated 
parametrization of $w$ both for thawing quintessence and for (a class of) freezing quintessence.

Apart from these motivations, it is also useful to derive $w(a)$ for more practical purposes 
because we do not need 
to solve the equation of motion directly for each potential. The idea has similarity, in spirit, 
with the slow-roll conditions: the existence of inflationary solutions reduces to 
simple conditions without having to solve the equation of motion directly. 

The paper is organized as follows: In Sec. 2, by perturbing the tracker equation, we 
derive the equation of state for tracker fields to all orders in $\Omp$. 
In Sec. 3, we derive the observational constraints on the parameter of the equation of state 
from Type Ia supernovae (SNIa) data and baryon acoustic oscillations (BAO). Sec. 4 is devoted to summary.


\section{Solving Tracker Equation}

We consider a flat universe consisting of background matter and scalar 
field dark energy $\phi$. 
The equation of motion of $\phi$ is 
\beqa
\ddot\phi+3H\dot\phi+V'(\phi)=0,
\label{eq:eom}
\eeqa
where $V'=\partial V/\partial\phi$. 
The equation of state $w$ is given by 
\beq
w=\frac{\dot\phi^2/2-V}{\dot\phi^2/2+V}.
\label{w}
\eeq
The equation of motion 
Eq.(\ref{eq:eom}) can be rewritten by using $w$ \cite{swz}:
\beq
\mp\frac{V'}{V}=\sqrt{\frac{3\ka^2(1+w)}{\Omp(a)}}
\left(1+\frac{\dot x}{6}\right),
\label{eomx}
\eeq
where the minus sign corresponds 
to $\dot\phi>0(V'<0)$ and the plus sign 
to the opposite, $\ka^2=8\pi G$,  $\Omp(a)$ is the density parameter of dark energy, 
$x=(1+w)/(1-w)$ and 
$\dot x\equiv {d\ln x}/{d\ln a}$.  
Tracker fields have nearly constant $w$ 
initially and eventually evolve toward $w=-1$.

\subsection{Tracker Solution}

Tracker fields 
have attractor-like solutions in the sense that a very wide range of
initial conditions rapidly converge to a common cosmic evolutionary 
track: the tracker solution \cite{swz}. Taking the derivative of Eq.(\ref{eomx}) with respect to 
$\phi$, we obtain the so-called tracker equation \cite{rubano,scherrer05,ktrac}
\begin{widetext}
\beqa
\Gamma-1\equiv \frac{VV''}{V'^2}-1=
\frac{w_B-w}{2(1+w)}-\frac{(1+w_B-2w)\dot x}{2(1+w)(6+\dot x)}
+\frac{3(w-w_B)\Omp(a)}{(1+w)(6+\dot x)}
-\frac{2\ddot x}{(1+w)(6+\dot x)^2},
\label{q-trac}
\eeqa
\end{widetext}
where $w_B$ is the equation of state of background matter 
 and $\ddot x\equiv d^2\ln x/d\ln a^2$. Equation (\ref{q-trac}) differs from the tracker equation 
in \cite{swz} where the term involving $\Omp$ is neglected which is essential in 
deriving the perturbation solution of $w$. 
 
Henceforth we consider the epoch after the matter-dominated era, so that we set $w_B=0$. 
Since during the matter-dominated epoch $\Omega_{\p}$ is negligible and $w$ becomes  
an almost constant for tracker fields, $w$ in that epoch is written in terms of $\Gamma$ as 
\cite{swz,ws}
\beq
\wzero=-\frac{2(\Gammazero-1)}{1+2(\Gammazero -1)},
\label{eosq}
\eeq
where the zero subscript in parentheses denotes the zeroth-order solution, 
neglecting the contribution of dark energy to the expansion rate. 


\subsection{Perturbing the Tracker Evolution}

In order to include the effect of finite $\Omp$, we treat it as a perturbation 
to the zeroth-order solution and then extrapolate the result to the situation where 
$\Omp$ is not so small when comparing the solution with the numerical solution. 

We define the perturbation to the zeroth-order  solution $\wzero$ to be 
$\delta w$, where $\delta w\sim \delta\phi/\phi\sim \order(\Omp)$ from Eq. (\ref{eomx}) \cite{ws}.
Keeping all terms of order $\Omp$ in Eq. (\ref{q-trac}), we obtain
\begin{widetext}
\beqa
a^2\frac{d^2\delta w}{da^2}+\frac{5-6\wzero}{2}a\frac{d \delta w}{da}+
\frac92 (1-\wzero)\delta w-\frac92 \wzero(1-\wzero^2)\Omp(a)
+9(1-\wzero^2)(1+\wzero)d\Gamma=0.
\label{deltaw}
\eeqa
\end{widetext}
We find that the analysis can be made simpler if $\Gamma={\rm constant}$ so that the last term in 
Eq.(\ref{deltaw}) is vanishing. 
This is the case for 
 inverse power-law potentials and for exponential potentials. 
In fact, for $V=M^4(M/\p)^{\alpha}$, 
$\Gamma=(1+\alpha)/\alpha$, 
and also for $V=M^4e^{\lambda\ka\p}$, $\Gamma=1$. 
We limit ourselves to the case when this holds so that we can  solve Eq. (\ref{deltaw}) 
without using $\delta\phi$. Note that this condition does not hold for 
a constant $w$ model \cite{lopez} and for an exponential of inverse power-law model: 
$V=M^4\exp(1/\ka\p)$ \cite{swz}.

By approximating $\Omp$ by the zeroth-order solution and expanding it 
in terms of the scale factor (or $\rho_{\phizero}/\rho_B$) as
\beqa
\Omp (a)&=&\frac{\Om_{\p 0}a^{-3\wzero}}{\Om_{\p 0}a^{-3\wzero}+(1-\Om_{\p 0})}\nonumber\\
&=&\sum_{n=1}^{\infty}(-1)^{n-1}\left(\frac{\Om_{\p 0}}{1-\Om_{\p 0}}\right)^na^{-3n\wzero}
\eeqa
we find, to all orders in  $\rho_{\phizero}/\rho_B$ or in $\Omp$, 
\begin{widetext}
\beqa
w(a)=\wzero+\delta w&=&\wzero+
\frac{(1-\wzero^2)\wzero}{1-2\wzero +4\wzero^2}\frac{\Om_{\p 0}}{1-\Om_{\p 0}}a^{-3\wzero}
-\frac{(1-\wzero^2)\wzero}{1-3\wzero +12\wzero^2}
\left(\frac{\Om_{\p 0}}{1-\Om_{\p 0}}\right)^2a^{-6\wzero} +\dots \nonumber\\
&=&\wzero+\sum_{n=1}^{\infty}\frac{(-1)^{n-1}\wzero(1-\wzero^2)}
{2n(n+1)\wzero^2-(n+1)\wzero+1}\left(\frac{\Om_{\p 0}}{1-\Om_{\p 0}}\right)^na^{-3n\wzero},\nonumber\\
&=&\wzero+\sum_{n=1}^{\infty}\frac{(-1)^{n-1}\wzero(1-\wzero^2)}
{2n(n+1)\wzero^2-(n+1)\wzero+1}\left(\frac{\Omp(a)}{1-\Omp(a)}\right)^n,
\label{eosfull}\\
&=&\wzero+
\frac{(1-\wzero^2)\wzero}{1-2\wzero +4\wzero^2}\Omp(a)
+\frac{(1-\wzero^2)\wzero^2(8\wzero-1)}{(1-2\wzero +4\wzero^2)(1-3\wzero +12\wzero^2)}
\Omp(a)^2
\nonumber\\
&&+\frac{2(1-\wzero^2)\wzero^3(4\wzero-1)(18\wzero+1)}
{(1-2\wzero +4\wzero^2)(1-3\wzero +12\wzero^2)(1-4\wzero +24\wzero^2)}
\Omp(a)^3+\dots \nonumber
\eeqa
\end{widetext}
Eq. (\ref{eosfull}) is our main result.\footnote{The infinite series in 
Eq. (\ref{eosfull}) can be written in terms of the 
 hypergeometric functions. } 
In the last equation, we have also expanded in terms of $\Omp(a)$ using 
$(\Omp(a)/(1-\Omp(a)))^n=\Omp(a)^n(\sum_{m=0}^{\infty}\Omp(a)^m)^n$.  
Up to the second order in $\Omp$, the solution becomes
\begin{widetext}
\beqa
w_2(a)=\wzero+
\frac{(1-\wzero^2)\wzero}{1-2\wzero +4\wzero^2}\Omp(a)
+\frac{(1-\wzero^2)\wzero^2(8\wzero-1)}{(1-2\wzero +4\wzero^2)(1-3\wzero +12\wzero^2)}
\Omp(a)^2.
\label{eos}
\eeqa
\end{widetext}
Note that $w(a)=-1$ if $\wzero=-1$ and hence the cosmological constant is contained in 
our $w(a)$. 
This $w(a)$ (or $w_2(a)$) agrees with the solution found in \cite{ws} (their Eq. (33)) up to the first order in 
$\Omp$: 
\beqa
w_{ws}(a)=\wzero+\frac{(1-\wzero^2)\wzero}{1-2\wzero +4\wzero^2}\Omp(a).
\label{eos1}
\eeqa

\begin{figure}
	\epsfig{file=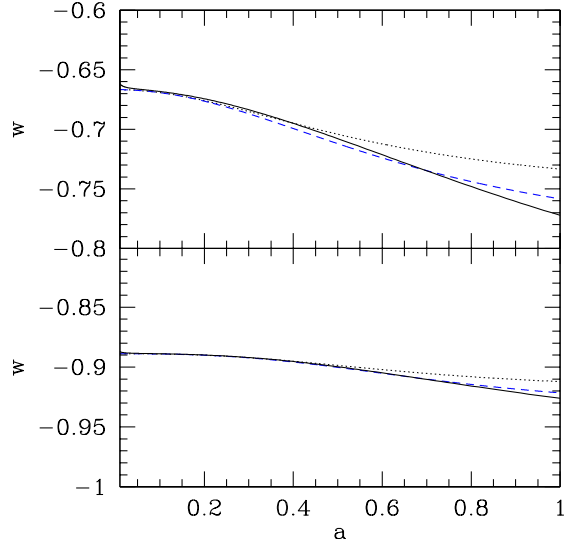,height=80mm}
	\caption
	{	\label{fig:eos} Evolution of $w(a)$ for inverse power-law potentials: 
$V=M^4(M/\p)^{\alpha}$. 
The upper plot corresponds to the $\alpha=1$ case and the lower one corresponds
	to the $\alpha=1/4$ case. The solid (black) lines denote the numerical result,  
the dashed (blue) lines denote the second order solution $w_2(a)$ given by \eqn{eos}, 
and the dotted lines denote 
the first order solution $w_{ws}(a)$ Eq. (\ref{eos1}).}
\end{figure}

We find the second order solution $w_2(a)$ Eq. (\ref{eos}) already agrees 
with the numerical solutions fairly well as shown in Fig. 1. 
For $\alpha \leq 1$ (or $\wzero \leq -2/3$), 
the fractional error of 
the equation of state between the numerical solutions and $w_2(a)$, Eq. (\ref{eos}), 
is less than 1.8\%, while the error can be as large as 5\% for  
$w_{ws}$.

\section{Observational Constraints on $w_{(0)}$}

We present  the observational constraints on 
the equation of state parameters $\wzero$. 
We use the second order equation of state $w_2(a)$ Eq. (\ref{eos}) for simplicity.\footnote{Including 
higher order terms does not affect the constraints.}

As observational data we consider the recent compilation of 397 SNIa, 
  called the Constitution Set with the light curve fitter SALT, by Hicken et al. \cite{hicken} and 
the measurements of BAO from the recent SDSS data \cite{baonew} 
which is now consistent with both earlier SDSS data \cite{bao} and 2dF data \cite{percival}. 
Uncertainties in the distance modulus of a supernova include uncertainties in 
light curve fitting parameters (the maximum magnitude, stretch parameter, color 
correction parameter) and due to the peculiar velocity (400 ${\rm km s^{-1}}$) 
as given in \cite{hicken}.

BAO  measurements from the SDSS data provide a constraint on the distance parameter 
 $A$ defined by
\beqa
A(z)=(\Om_mH_0^2)^{1/2}\left(\frac{1}{H(z)z^2}\right)^{1/3}\left(\int^z_0\frac{dz'}{H(z')}\right)^{2/3}
\eeqa
to be $A(z=0.35)=0.493\pm0.017$ \cite{baonew}.

\begin{figure}
\epsfig{file=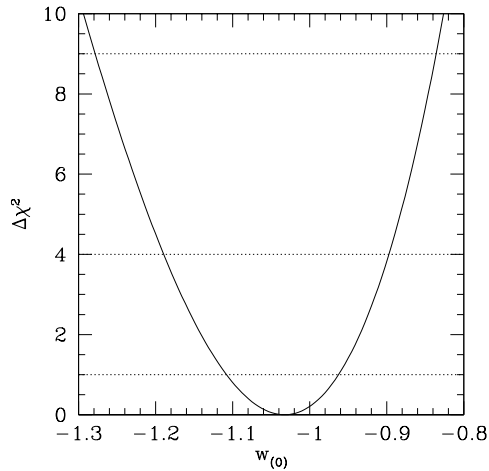,height=70mm}
\caption{$\Delta \chi^2$ as a function of $\wzero$. }	
\label{sndata}
\end{figure}

The $\chi^2$ curve normalized by its minimum, $\Delta\chi^2=\chi^2-\chi^2_{min}$, 
calculated from SNIa and BAO is shown in Fig. \ref{sndata}. 
We marginalize over $\Om_m$ to calculate the curve. 
The allowed range of $\wzero$ is narrow: $-1.11< \wzero< -0.96 (1 \sigma)$, 
$-1.19< \wzero< -0.90 (2 \sigma)$, $-1.28< \wzero< -0.84 (3 \sigma)$.

\section{Summary}

We have derived the equation of state for a class of freezing quintessence models 
called tracker fields whose equation state is nearly constant during the matter-dominated era. 
By solving the tracker equation perturbatively, we could 
derive a useful approximated solution to the equation of state for tracker fields (Eq. (\ref{eosfull})). 
The solutions agree with the numerical solutions quite accurately. 

Our solution is also useful for pragmatic purposes in that one only have to use 
our equation of state without solving the scalar field of motion numerically for any 
power index $\alpha$. 

Applying the solution of $w(a)$ truncated to the second order in $\Omp$, $w_2(a)$,  
to SNIa data and BAO, we find that the parameter $\wzero$, 
which parameterizes the equation of state during the matter-dominated era, 
is constrained to lie near $-1$ and hence the cosmological constant limit of 
these models is consistent with the current data. 

Combining with our previous results on thawing models \cite{ds,chiba,cds}, 
we now have three parameters 
($w_0,K,\wzero$) for the equation of state both for thawing models and for (a class of) freezing models:
\beqa
1+w(a)=\left\{
\begin{array}{lr}
(1+w_0)a^{3(K-1)}\left(\frac{(K-F(a))(F(a)+1)^K+(K+F(a))(F(a)-1)^K}
{(K-\Om_{\p 0}^{-1/2})(\Om_{\p 0}^{-1/2}+1)^K+(K+\Om_{\p 0}^{-1/2})(\Om_{\p 0}^{-1/2}-1)^K}\right)^2
   & \mbox{(thawing quintessence)},\\
1+ \wzero+\sum_{n=1}^{\infty}\frac{(-1)^{n-1}\wzero(1-\wzero^2)}
{2n(n+1)\wzero^2-(n+1)\wzero+1}\left(\frac{\Omp(a)}{1-\Omp(a)}\right)^n  & \mbox{(freezing quintessence)},
\end{array}
\right.
\eeqa 
where $F(a)=\sqrt{1+(\Om_{\p0}^{-1}-1)a^{-3}}$. In this dark energy theory space, 
 dark energy is close to the cosmological constant, which corresponds to $w_0=-1$ 
irrespective of $K$ and $\wzero=-1$, to the extent such that:\footnote{The constraint on $w_0$ is 
recalculated using the recent SDSS data \cite{baonew}, which only slightly shifts toward 
negative $w_0$.} 
$-1.14<w_0<-0.92, -1.11<\wzero<-0.96$ and no constraint on $K$.

\section*{Acknowledgments}
The author would like to thank Masahide Yamaguchi for useful communications. 
This work was supported in part by a Grant-in-Aid for Scientific Research
from JSPS (No.\,20540280)
and from MEXT (No.\,20040006) and in part by Nihon University. 
Some of the numerical computations were 
performed at YITP at Kyoto University.




\end{document}